\documentclass[%
prl,
reprint,
superscriptaddress,
showpacs,
 amsmath,amssymb,
 aps,
floatfix,
]{revtex4-1}
\usepackage[bottom]{footmisc}
\usepackage{graphicx}
\usepackage{dcolumn}
\usepackage{bm}
\usepackage{color}
\usepackage[caption = false]{subfig}
\usepackage{floatrow}
\usepackage[mathlines]{lineno}

\newcommand{\tauliq}{\tau_\mathrm{liq}}
\newcommand{\taunet}{\tau_\mathrm{net}}
\newcommand{\Esw}{E_\mathrm{sw}} 

\begin{document}


\title{Dynamics of vitrimers: defects as a highway to stress relaxation}

\author{Simone Ciarella}
 \affiliation{%
 Department of Applied Physics, Eindhoven University of Technology, Postbus 513, NL-5600 MB Eindhoven, Netherlands
}%
\author{Francesco Sciortino}%
 \affiliation{%
 Department of Physics and CNR-ISC, Sapienza Universit\'a� di Roma, Piazzale Aldo Moro 2, I-00185 Roma, Italy
}%
\author{Wouter G. Ellenbroek}%
 \affiliation{%
 Department of Applied Physics, Eindhoven University of Technology, Postbus 513, NL-5600 MB Eindhoven, Netherlands
}%
\affiliation{Institute for Complex Molecular Systems, Eindhoven University of Technology, Postbus 513, NL-5600MB Eindhoven, The Netherlands}





\date{\today}

\begin{abstract}
We propose  a coarse-grained model to investigate stress relaxation
in star-polymer networks induced by dynamic bond exchange processes.
We show how  the swapping mechanism, once activated,
allows the network to reconfigure, exploring distinct topological configurations,
all of them characterised by complete extent of reaction.  Our results
reveal the important role played by topological defects in mediating the
exchange reaction and speeding up stress relaxation.
The model provides a  representation of the dynamics in
vitrimers, a new class of polymers characterized by bond swap mechanisms which preserve the total number of bonds, as well as in other bond-exchange materials.
\end{abstract}

\maketitle

Vitrimers, an exciting new class of polymer networks, are unique in their
ability to interpolate between the two  conventional classes of polymer,
thermoplastics and thermosets~\cite{rubinstein2003polymer}. The first can be
reshaped at will, but are sensitive to being weakened by contact with solvents,
while the latter are insoluble but cannot be reshaped after the cross-linking
process.  In vitrimers,  a connectivity-preserving bond exchange
mechanism~\cite{kloxin2013covalent,denissen2016vitrimers,montarnal2011silica,capelot2012metal}
with well-controlled exchange rate makes the cross-links dynamic.  At low
rates, they perform like thermosets, while at high rates, they are malleable
like thermoplastics.
Unlike permanently cross-linked elastomers or gels,
these bond swaps allow vitrimers to release internal stresses without losing shape.
Their versatility shines particularly in
smartly designed materials, where bond swapping provides a welding
strategy~\cite{chabert2016epoxy}, or responsiveness to light, pH, voltage,
metal ions, redox chemicals and mechanical
stimuli~\cite{chen2017oligoaniline,yang2016graphene,ruiz2016epoxy}. 

The unusual molecular interaction in vitrimers renders
current theories of polymer performance of limited use. Neither a
conventional static model nor a fully dynamic one can coherently address the
exchange dynamics.
At the atomistic level, exchange reactions can be effectively modeled using
reactive force fields~\cite{van2001reaxff}. This gives a detailed picture of
a single exchange event, but does not provide large enough time and length scales to
assess macroscopic properties. To get to macroscopic scales, a coarse-grained model that 
captures the network-topology aspects of the exchange reactions is needed.

In recent years, scientists have developed different numerical models to study
exchange
materials~\cite{wittmer2016shear,stukalin2013self,smallenburg2013patchy},
embedding Monte Carlo hopping moves into hybrid molecular dynamics or Monte
Carlo (MD,MC) simulations to reproduce bond swaps.
In this letter, we study a vitrimer model consisting of associative star-polymers
using a three-body potential to reproduce bond exchange dynamics with a controllable
rate~\cite{sciortino2017three}, avoiding the need for hybrid features.
Using molecular dynamics simulations to obtain
the stress relaxation modulus, we verify the expected transition from
solid-like to liquid-like long-time behavior 
upon increasing the bond exchange rate \cite{snijkers2017curing}.  More
importantly, we uncover a dramatic difference in stress relaxation that arises
from the molecular topology. In close connection to recent work that
demonstrated how loops affect equilibrium elastic
properties~\cite{zhong2016quantifying}, we show that networks made from
building blocks that allow loop formation via bond exchange relax stresses much
faster than systems made from loop-preventing building blocks, even when the
bond exchange rates are the same.  Thus, the slow relaxations that
characterize swapping vitrimers~\cite{rovigatti2018self} can be controlled not
only through the swap rate, but also through defect formation. In this sense,
loop defects serve as highways to stress relaxation,
giving faster self-healing and better malleability and
recyclability~\cite{montarnal2011silica,stukalin2013self}. Combining this
effect with an accurate choice of the network topology we can imagine to
synthesize a material which is not only stable as thermosets and malleable as
thermoplastics, but also tougher than either, because it can relax stresses in
a controllable way, while having improved structural integrity compared to
materials toughened via other mechanisms such as fully reversible
crosslinking~\cite{kean2014invisible} or sacrifical
bonds~\cite{ducrot2014toughening}.

\begin{figure*}[!bth]
\includegraphics[width=\textwidth]{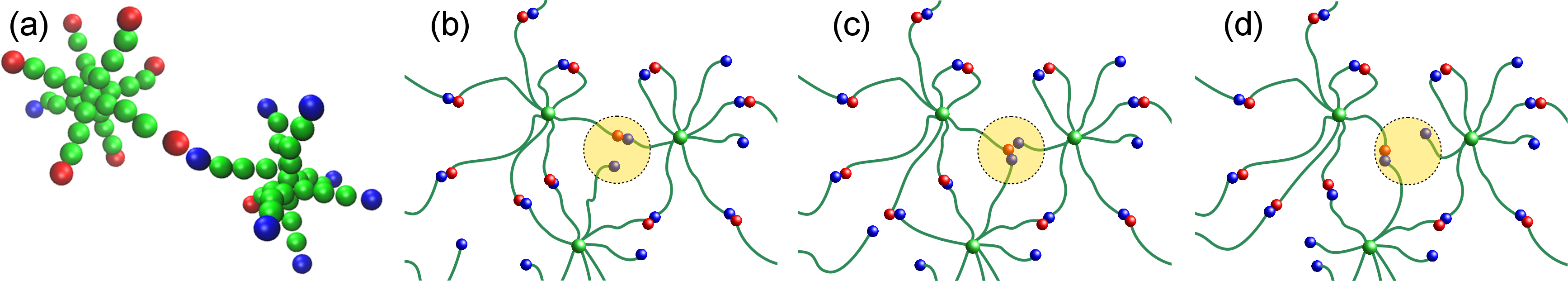}
\caption{ (a) Our star-shaped monomers forming a swappable covalent bond (red-blue). (b)-(d) Sketch of a swap event: red-blue bonds can swap, while green beads are permanent links
(the centers of the stars). The swap reaction modifies the topology of the
network. Its rate is catalytically controlled in
real systems, modeled here by tuning the energy barrier.
In states
(b) and (d) there is only a two-body energy term ($\approx -\epsilon$) due to the highlighted bond. In state (c)  there are two two-body energy terms ($\approx -2\epsilon$) and one three-body contribution ( $\approx +\lambda \epsilon$).
If  $\lambda=1$,  the three-body term compensates exactly the
formation energy of the second bond effectively flattening out the energy
barrier.
}
\label{f_intro}
\end{figure*}

\emph{Modeling Vitrimers---}%
We focus our simulations on networks built from binary mixtures of eight-arm
star polymers. Each arm terminates with a reactive site  which can be of two
different types, labeled red and blue (see Fig.~\ref{f_intro}(a)).  This
effectively captures what happens in vitrimers that rely on covalent association of
two different moieties, via e.g.\ ester 
bonds~\cite{montarnal2011silica,capelot2012metal}, in which case 
the end types represents carboxyl and hydroxyl groups, respectively.
Star-shaped monomers are widely used polymeric building
blocks, for e.g.\ dendrimers and tetra-PEG hydrogels~\cite{sakai2008tetrapeg,matsunaga2009structure}.
They are a versatile basis for
covalent adaptive networks, with controlled connectivity
and architecture~\cite{kloxin2013covalent}.

We coarse grain the star-polymer as a sequence of beads and
harmonic springs~\cite{kremer1990dynamics}
with a rest length of $1\,$nm which is our unit of length. 
The beads, shown in green in Fig.~\ref{f_intro}(a), thus represent Kuhn segments consisting of
roughly 8 carbon atoms. Masses, energies and times are expressed in units
$[m]=100\,$u, $[E]=k_\mathrm{B}\cdot 300\,$K, and $[\tau]=6.33\,$ps, respectively. All pairs of beads
interact via a purely repulsive WCA potential ($\sigma=1\,$nm)~\cite{wca1971}.

Modelling swappable covalent bonds using potentials requires care.
They must enforce single red-blue bonds without clustering, contain
a parameter to tune the swap rate, and the bonds they provide
must be thermally stable.
To this end, we
use a combination of two-body and three-body interactions as proposed
in Ref.~\cite{sciortino2017three}. The two-body term is a generalized Lennard-Jones 
potential acting only between red-blue pairs,
\begin{equation}
v_{ij}\left(\vec{r}_{ij}\right)=
 4 \epsilon \left[ \left( \dfrac{ \sigma}{r_{ij}} \right) ^{20}- \left( \dfrac{ \sigma}{r_{ij}} \right)^{10} \right] \qquad r<r_\mathrm{cut}~.
\end{equation}
With $\sigma=0.5\,$nm and $\epsilon=100\,k_\mathrm{B}T$, 
the  $v^{\left( 2b \right)}_{ij}$ provide a covalent-like bond that
is stable against thermal fluctuations. We fix $r_\mathrm{cut}=2.5\sigma$.
The 3-body term is rewritten in terms of how the interaction between particles $i$ and $j$ is affected by the presence of other particles $k$ that are within range of particle $i$,
\begin{equation}
\label{eqv3}
v^{\left( 3b \right)}_{ijk}=\lambda \epsilon\,\hat{v}^{ \left( 2b \right)}_{ij}\left(\vec{r}_{ij}\right) \cdot \hat{v}^{ \left( 2b \right)}_{ik}\left(\vec{r}_{ik}\right)~,
\end{equation}
where $\lambda \geq 1 $ is the three-body scaling parameter and $\hat{v}^{ \left( 2b \right)}_{ij}$ is defined as 
\begin{equation}
\hat{v}^{ \left( 2b \right)}_{ij}\left(\vec{r}_{ij}\right) = 
\begin{cases}
& 1 \qquad \qquad \; \; \qquad r\le r_{min}\\
& - \dfrac{v_{ij}\left(\vec{r}_{ij}\right)}{\epsilon} \qquad r > r_{min}~.\\
\end{cases}
\end{equation}
Because it is formulated in terms of the attractive part of the two-body term,
this three-body potential compensates the pair energy that would be gained by 
two simultaneous red-blue bonds so that all intermediate states
encountered during a swap event are similar in potential energy. This flat energy landscape
is the defining feature of the method, as illustrated in Fig.~\ref{f_intro}(b-d).
The three-body term automatically enforces the single-bond per reactive site since
it gives a strong repulsion when more than 3 reactive sites are close.
The parameter $\lambda$ sets the energy barrier for a swap
rearrangement $\Delta \Esw$. To a first approximation $ \beta \Delta \Esw
 \equiv \beta \epsilon(\lambda-1)= 100(\lambda-1)$.

While three-body interactions are generally expensive in simulations,
Eq.~(\ref{eqv3}) requires only small additional numerical effort compared to a
standard two body potential, because it is a combination of the existing
two-body terms.

\emph{Numerical approach---}%
We use the Hoomd-blue package~\cite{anderson2008general,glaser2015strong} to
do molecular dynamics simulations on GPUs. For the three-body potential,
we developed a Hoomd-blue module named ``RevCross''.
This implementation allows us to 
gather sufficient statistics for evaluating the stress relaxation in systems of $N\approx50000$ beads ($\approx 1500$ star polymers).
To provide a reservoir of
open endings that can initiate a swap event we use
a non-stoichiometric mixture of different star endings, following the chemistry
behind vitrimers. We exploit two different mixtures to assess the role of defects (loops) in the stress relaxation.


We focus on the type of defects known as primary loops, in which two endings of the same star are bonded together.
These are the most important for the static
elastic properties~\cite{zhong2016quantifying}. First, we employ a defect-free mixture (DFM)
composed of $N_A=900$ 8-arm star polymers
whose endings are only type A and $N_B=600$ stars with only B-type ends.
Since A--B bonds are allowed, primary loops are prevented. Later, we present
results on a defect-allowing mixture (DAM) which contains $N_A=950$ stars with seven A-type
endings and a single B-type ending, and $N_B=550$ stars with the numbers reversed.
These values of $N_A$ and $N_B$ make the total number of red and blue beads
identical in both mixtures.  Since the red-blue bonds are much stronger than $k_\mathrm{B}T$, all 4800 B-type ends
will form a bond, leaving 2400 free A-type ends 
available to initiate swap events.
The large number of arms is used in order to have a network that behaves like a solid without
applying any (osmotic) stretching. 
Both networks are equilibrated in periodic cubic boxes of size $L=40\,$nm, corresponding to a packing fraction $\phi\approx0.3$.
This corresponds to 2.2 times the overlap concentration,
so the stars can easily form a network, but it is low enough to avoid any glassy dynamics. In the supplementary information,
we demonstrate that indeed there is no caging or segmental slowing down at this density, so that the polymer arms
are mobile enough to initiate bond swaps~\cite{supplement}.
For both DFM and DAM mixtures, we generate $m=100$ independent network topologies.

\emph{Stress relaxation---}%
We perform stress relaxation calculations. Rather than doing out of
equilibrium MD calculating the stress $\sigma(t)$ after a step strain,
we exploit the widely used autocorrelation method
\begin{equation}
G(t)\approx C(t)\equiv \frac{V}{k_\mathrm{B}T} \left< \overline{\sigma_{yz}(t) \sigma_{yz}(0)}\right>
\end{equation} 
in the $nVT$ ensemble where we imposed the number of stars
 $n$, the volume $V$,
and the temperature $T$. The bar and brackets denote averaging over time and ensemble, respectively. To calculate the instantaneous stress $\sigma(t)$ we have to add terms that arise from the three-body potential to the standard (pair-based) virial expression.
In the SI~\cite{supplement}, we derive these terms from the thermodynamic definition of stress.

The stress autocorrelation function is often assumed to be equal to the stress relaxation $G(t)$, but it was recently pointed out that the equality holds only in liquids~\cite{wittmer2015shear,kriuchevskyi2017numerical}. Still, for self-assembled networks, $C(t)$ \textit{on average} converges to $G(t)$~\cite{wittmer2016shear}. 
The correct way to define the stress relaxation would be
\begin{equation}
\label{eq:G}
G(t)=\begin{cases} C(t),& \mbox{liquids}\\C(t)+G_\mathrm{eq}-C_{\infty}, & \mbox{solids}\end{cases}
\end{equation}
where $G_\mathrm{eq}$ is the shear modulus and $C_\infty$ is the long-time asymptote of $C(t)$ (so $C_\infty\propto\langle\bar{\sigma}\rangle^2$).
Thus, the stress autocorrelation function $C(t)$ and the stress relaxation modulus
$G(t)$ always coincide in the liquid phase,
but when the system rigidifies, $C(t)$ shifts from $G(t)$ by a constant~\cite{wittmer2015shear}.
While they become identical only in thermodynamic
limit~\cite{wittmer2016shear}, we can distinguish a solid from a
liquid using the limiting behavior of $C(t)$, even in a finite ensemble. The reason this works
is that the only way to have
$C_{\infty}=0$ is when $\bar{\sigma}=0$ for every configuration, which happens
only for liquids. In the following we will simply use $C(t)$.



\begin{figure}[!pbt]
\includegraphics[width=\columnwidth]{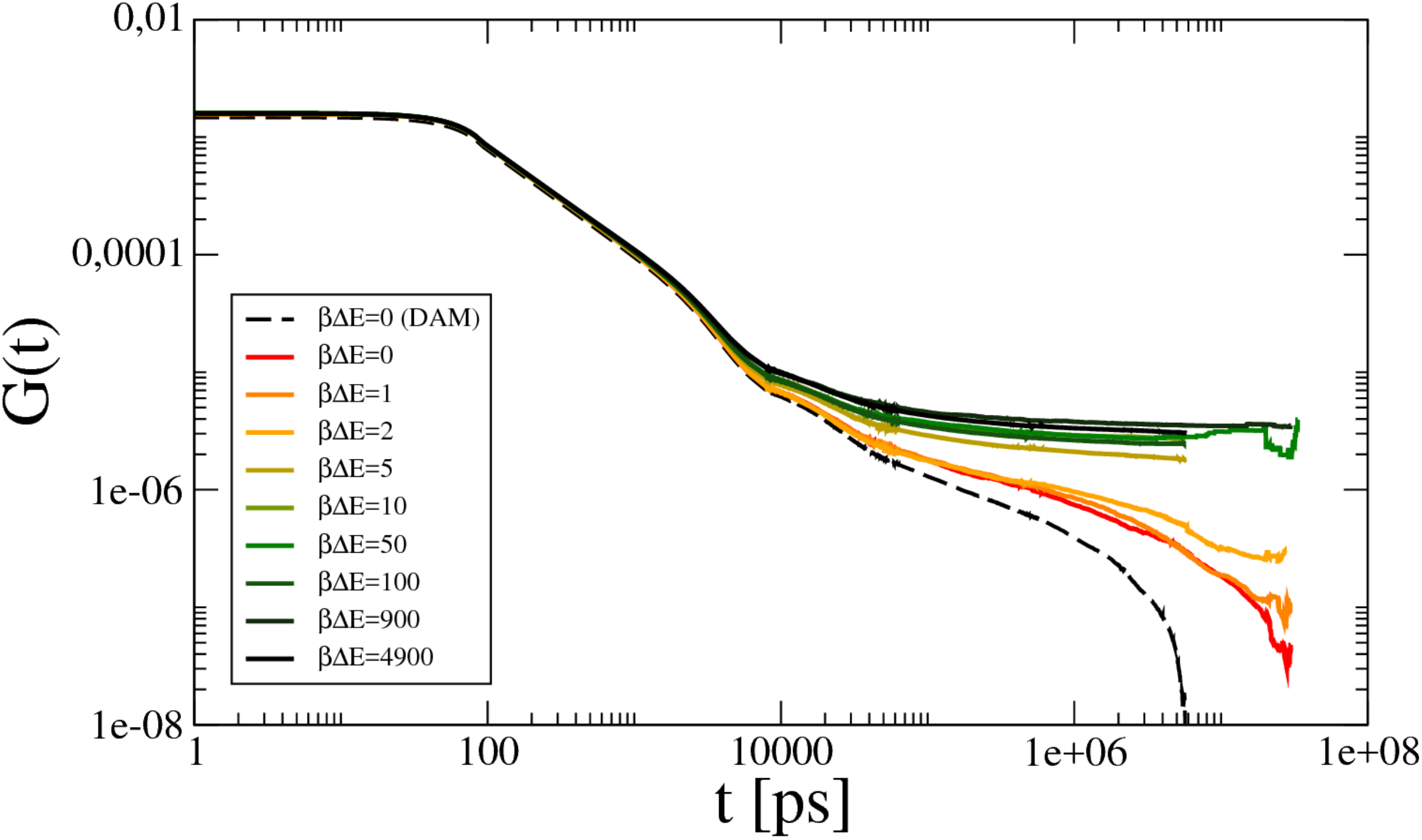}
\includegraphics[width=\columnwidth]{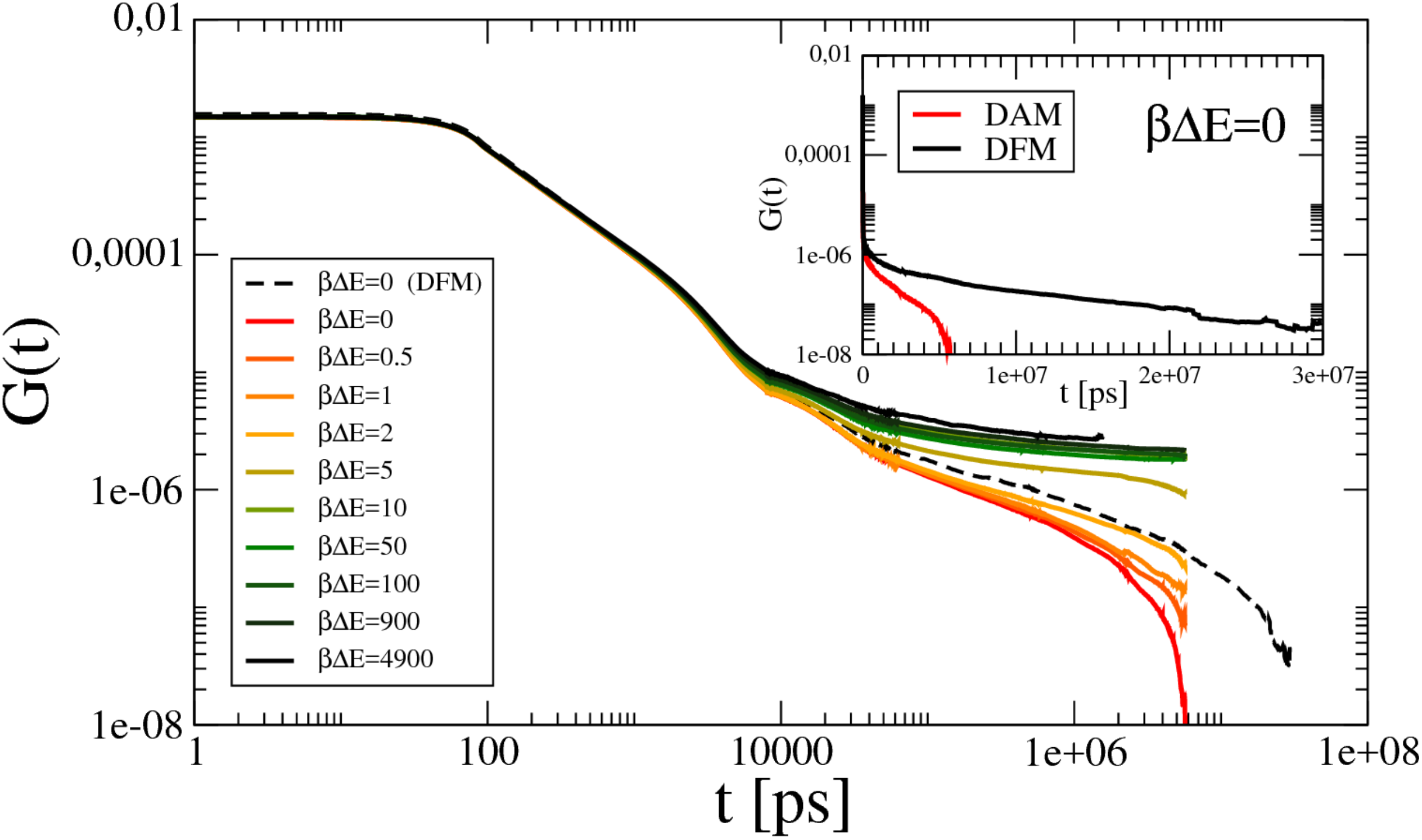}
\caption{(a) Stress relaxation for the Defect Free Mixture (DFM), for a range of swap barrier values (DAM data shown as dashed line for comparison). 
(b) Stress relaxation for the Defect-Allowing Mixture (DAM), with DFM data shown as dashed line for comparison.
After the first regime of relaxation due to chain rearrangement, a solid plateau is approached. For low enough energy barriers, swap
rearrangements trigger a second relaxation. This network relaxation is an order of magnitude faster with defects than without. Inset: DAM-DFM comparison on a linear time axis.}
\label{fig_rel}
\end{figure}
\emph{Swap-driven transition without loops---}%
The stress relaxation for the DFM system is reported in Fig.~\ref{fig_rel}a. 
We define $\taunet $ as roughly the time that it takes for a solid network to reach its elastic plateau, $\taunet \approx 5\,$ns.
At short times ($t<\taunet$), the stress relaxation is dominated by the Rouse modes of the chains~\cite{rouse1953theory}. We refrain from
fitting a power law to this regime because the arms of the star polymers are too short to make this feasible.
All curves coincide until this timescale because swapping is slow and the network topology is
essentially fixed. Then the gel starts sustaining the stress and a plateau in $G(t)$ appears.
If the swap move has a large energy barrier ($\beta \Delta \Esw>50$), the topology remains fixed and the plateau extends beyond
times reachable by simulation.
If instead the gel rearranges through bond swap moves, we observe a second
relaxation, the hallmark of transient networks~\cite{wittmer2016shear,stukalin2013self,smallenburg2013patchy,bomboi2016re}.
We conclude that 
when there is no activation barrier,  
swaps make DFM liquid at $\tauliq^{DFM} \approx 20\,\mu$s, where we picked $G(\tauliq)/G(0)\equiv10^{-4}$.

\emph{Swap-driven transition with loops---}%
The  stress relaxation for the DAM system is 
 reported in Fig.~\ref{fig_rel}b. After chain
relaxation, the solid plateau is approached only by the fixed networks, while
the swapping ones keep relaxing all of their stress. In this mixture, bond swapping contributes to stress relaxation on shorter
time scales. In marked contrast with the system without loop defects, the final stress
relaxation is now ten times faster, 
 $\tauliq^{DAM} \approx 0.1 \cdot \tauliq^{DFM} \approx 2\,\mu$s.
The swapping gel with defects behaves essentially like a viscous liquid.

To rule out that the faster stress relaxation is caused by structural
quantities unrelated to loop defects, we verified that all 2400 possible bonds
formed in both mixtures, and that the number of swap events is similar in both. 
Finally, we checked that the DAM mixture indeed formed loop defects, and found
that typically between 1/5 and 1/3 of all bonds in this mixture are primary loops,
connecting two arms of the same star.

\begin{figure}[!pt]
\includegraphics[width=6.5cm]{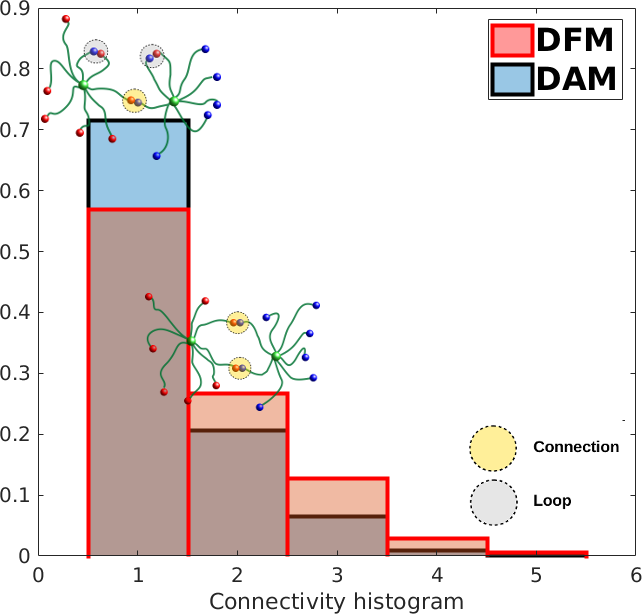}
\caption{Histogram of the number of connections between connected stars, for both mixtures.
In the DAM (blue), the presence of loops causes the average number of connections between stars to be lower. This reduces redundancy
and thereby increases the chance that a single swap even will disconnect two stars, which speeds up stress relaxation.}
\label{fig_histo}
\end{figure}

The conclusion is that the fast stress relaxation of the defect-allowing
mixture is caused by ``defected'' configurations.
Whenever an intra-star bond
that was carrying stress is swapped with an arm on the same star to form a
defect, it ceases to support stress, giving rise to dissipation.
This consistently leads to fewer redundant connections between starts, as we show in
Fig.~\ref{fig_histo}, which in turn makes the fraction of swap events that actually
detach two stars even larger, leading to more relaxation per event.
Given that the swap rate is similar for the two mixtures,
this means the defects act as a highway to stress relaxation.

\begin{figure}[!pt]
\includegraphics[width=\columnwidth]{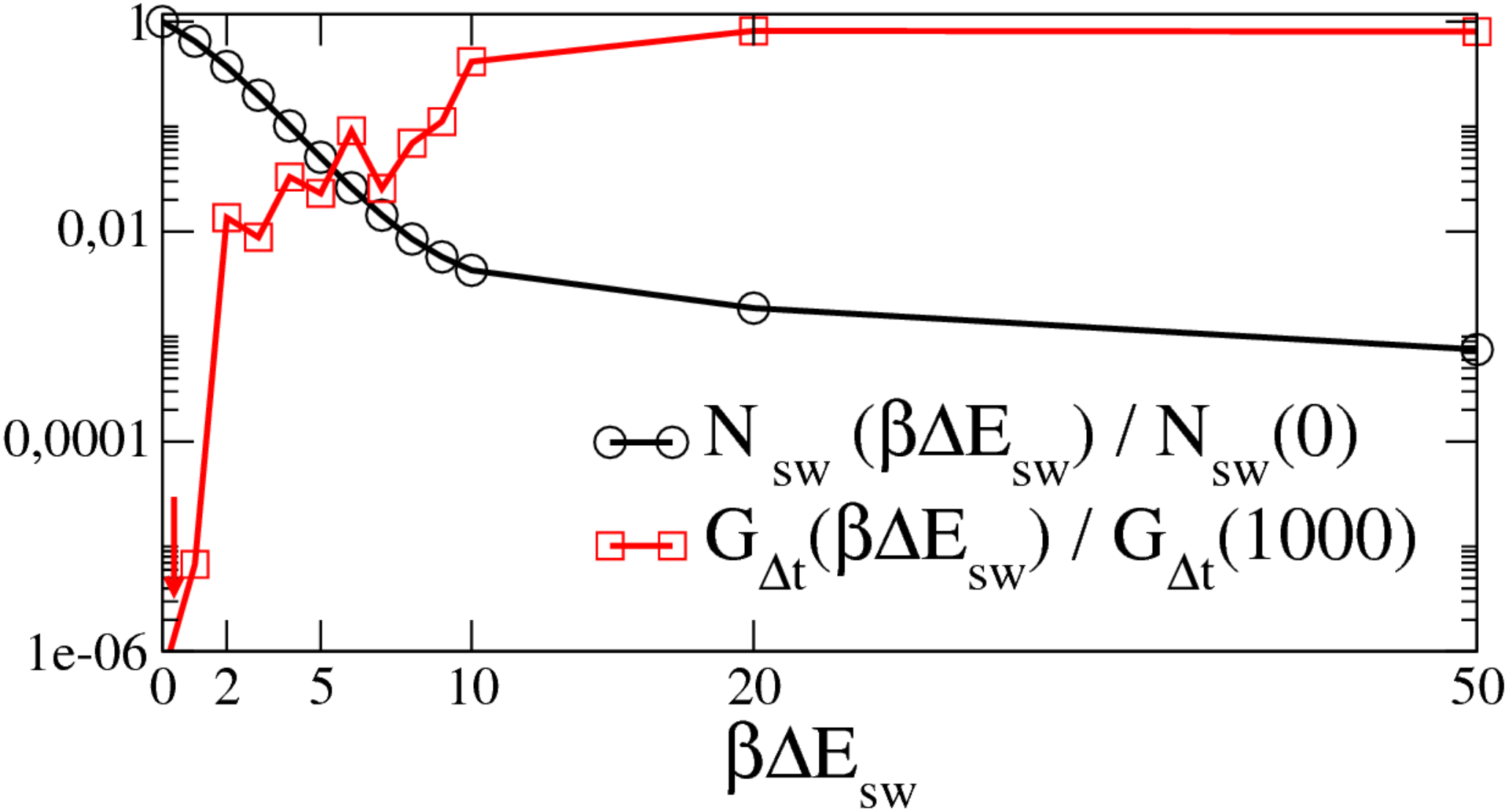}
\caption{ 
Normalized number of swaps (black) and stress relaxation (red) as a function of the swap energy barrier $\beta \Delta \Esw$, for the DAM system.
The fluid-solid transition happens when the barrier is $10 k_BT$.}
\label{fig_sw}
\end{figure}
\begin{figure}[!pt]
\includegraphics[width=\columnwidth]{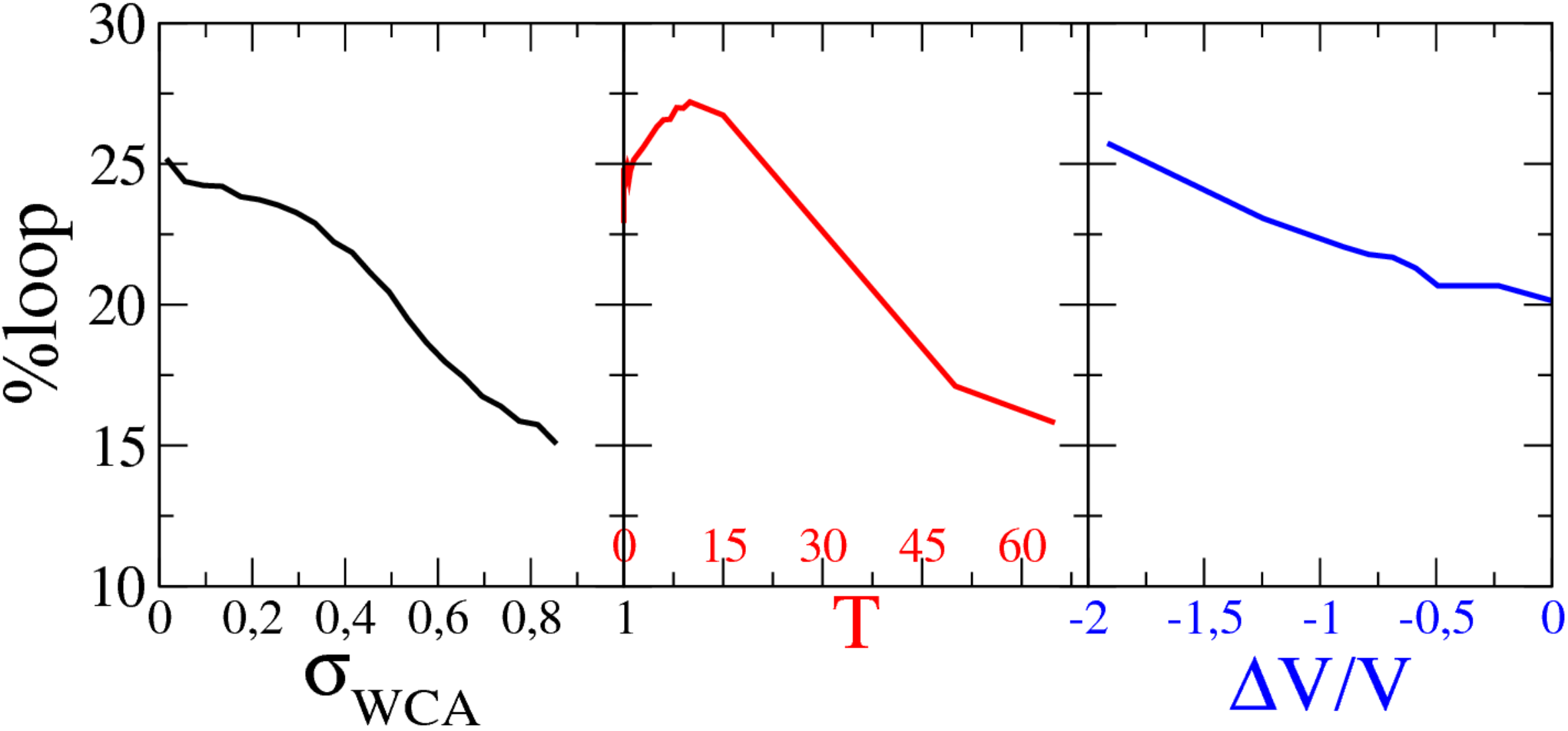}
\caption{We can control the number of loop defects by inflating the monomers (black), increasing temperature (red), or volumetric compression (blue).}
\label{fig_loops}
\end{figure}

\emph{Discussion---}%
We have shown the effectiveness of MD simulations in the study of bond swapping
systems. The employed three-body potential turns bond-swap events into a
continuous process: Free binding sites approach an
existing bond via a tunable barrier,
and after the exchange the unbound partner leaves via the
same pathway. The computational effort of evaluating a three-body
interaction is partly mitigated by defining it in terms of 
pair forces which had to be computed anyway. Compared to Monte Carlo
moves for bond exchanges, our approach has the advantage that
dependence of exchange probability on physical parameters such as the bond
force arises naturally and does not have to be manually added into an
acceptance criterion.

Fig.~\ref{fig_sw} shows the swap rate decreases by two orders of magnitude when
$\beta \Delta \Esw$ is increased from zero to six, correlated with the rise of
the elastic plateau in the stress relaxation modulus $G(t)$. Thus, $\beta
\Delta \Esw$ controls the solid-liquid transition or topological glass
transition in the same way catalyst concentration or temperature do this in
experiments.

Applying this method to two different mixtures, we demonstrate the importance
of topological considerations for stress relaxation, extending recent insights
into the effect of loop-like defects on static elastic properties~\cite{zhong2016quantifying,
wang2017oddeven}. With the same number of swapping events, our primary-loop-free mixture of star polymers
relaxes stress much more slowly than the loopy mixture of otherwise similar star polymers.

We stress the peculiar role for doubly connected stars in these networks: For the static modulus,
there is little difference between a single or double polymer bridge between two star centers, as
each bridge in a second-order loop is about half as effective as a single bridge~\cite{zhong2016quantifying}.
When it comes to relaxing stresses via swapping, however, the doubly connected stars
contribute much more slowly since the force between them will only be relaxed away after both bonds have
undergone a swap.

Interestingly, we found that we can exert some control over how many loops are
formed in the defect-allowing mixture by way of excluded volume interactions, temperature and deformation 
as we show in Fig.~\ref{fig_loops}. This observation may open up ways to enhance control over
the elastic properties of networks, which is a topic that deserves further
study in the context of vitrimers and other bond-swapping materials. 

While in standard cross-linked networks the density of loops can be controlled through the 
reaction protocol~\cite{wang2017oddeven}, we speculate that the state at which a vitrimer is allowed to equilibrate
topologically, determines the number of defects at equilibrium. If we then
do a rapid temperature quench to a regime where bond exchange is inhibited,
the topology is frozen in a state with a controlled average number of defects.


In summary, our results suggest that novel vitrimer systems
can be designed explicitly considering defects as a means to control mechanical properties.  
This class of polymeric materials will then, on top of their recyclability and
their remarkable ability to recover their initial properties after remolding,
also become mechanically tunable.

\emph{Acknowledgments---}We are grateful to Hans Heuts and Cornelis Storm for helpful comments.

\providecommand{\noopsort}[1]{}\providecommand{\singleletter}[1]{#1}

\onecolumngrid
\appendix
\clearpage
\textbf{SUPPLEMENTAL INFORMATION included as appendices}
\section{APPENDIX A --- Two glass transitions}
\begin{figure}
\floatbox[{\capbeside\thisfloatsetup{capbesideposition={right,center},capbesidewidth=5cm}}]{figure}[\FBwidth]
{\caption{Long time limit of the stress relaxation function (DAM) as a function of density. In the inset we plot the bond de-correlation function at $\tauliq\approx1\,\mu$s. Above the glass transition bond swaps are blocked.}\label{fig_glass}}
{\includegraphics[height=6cm]{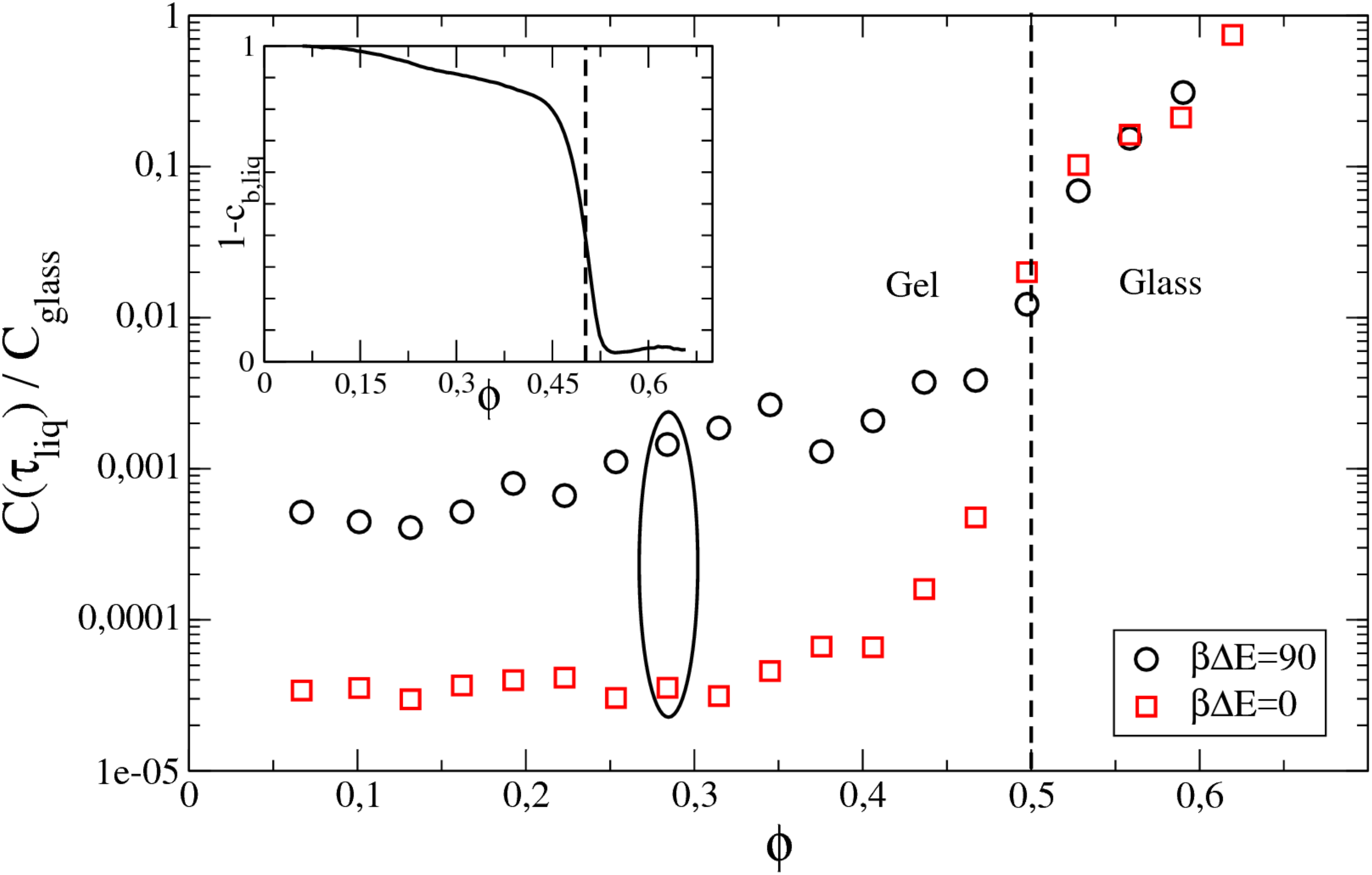}}
\end{figure}
Vitrimers are usually described as systems with two glass transitions. The
standard glass transition describes how chains or parts thereof become
kinetically arrested upon lowering temperature or increasing density, and
happens in melts and crosslinked networks alike. The second glass transition is
specific to vitrimers and describes topological arrest. When conditions allow
bond swapping, vitrimers are able to rearrange their network structure following a
valley-rich path in a rough energy landscape. Without swapping, the network is
stuck in a particular topology (polymer architecture). In experiments this
topological glass transition is controlled through temperature and catalyst concentration~\cite{capelot2012metal,denissen2015vinylogous}. In
our model, correspondingly, through the activation energy for swap $\beta\Delta \Esw$. 

Hence at high temperatures, vitrimers can flow like viscoelastic liquids~\cite{denissen2016vitrimers,snijkers2017curing},
while they become solid at low temperature or high density through a standard glass transition
or at low swap rate through a topological one.
The interesting physics of vitrimers is related to their unique topological glass
transition, which being reversible and easily controllable in experiments,
opens up to a lot of technological and industrial applications~\cite{chen2017oligoaniline,yang2016graphene,ruiz2016epoxy,chabert2016epoxy}.

To demonstrate that we are looking only at the effects of the topological glass
transition we estimated the density at which the standard kinetic glass transition
happens. We plot in Fig.~\ref{fig_glass} the long time limit of $C(t)$ during a compression/decompression
protocol. The protocol consists in the equilibration of the same $100$ DAM networks with 
$\beta \Delta \Esw=0$ at different volumes for a time $t\approx10\tauliq$, and then a measure of the stress 
relaxation $C(\tauliq)$~\cite{tauliq_mes} for both viscous liquids ($\beta \Delta \Esw=0$) and topologically 
frozen gels ($\beta \Delta \Esw=90$). We express our results as a function of effective hard sphere packing 
fraction, defining $V_\mathrm{WCA}(\sigma_\mathrm{HS})=k_\mathrm{B}T$.
We go from $\phi\approx 0.3$ to $\phi_\mathrm{max}\approx 0.61$, which is the point
where the equilibration time becomes $t_\mathrm{eq}\gg10\tauliq$. 
While decompressing we go down to $\phi\approx 0.05$.
The dotted line in Fig.~\ref{fig_glass} corresponds to $\phi=0.5$.
We can see from the plot that the system behaves the same
up until crossing $\phi=0.5$  where the slowdown starts and the caging effect
prevents a full relaxation. After this point in fact, even the $\beta \Delta \Esw=0$ 
networks behave as solids. We then measure the bond autocorrelation function
$c_b(t)$, which we use to plot the long time limit of the de-correlation function
$1-c_b(\tauliq)$ in the inset of Fig.~\ref{fig_glass}. As hypothesized, above the
glass transition bond swaps have no effect. The choice of $\phi\approx0.3$ that we do in the main text,
corresponds to the points in the black ellipse, where the density is low enough
to exclude any interference of the glass transition, so that bond-swap-induced structural rearrangements
can contribute to stress relaxation.

\section{APPENDIX B --- Pressure and stress in the presence of three body interactions}
\begin{figure}
\floatbox[{\capbeside\thisfloatsetup{capbesideposition={left,center},capbesidewidth=4cm}}]{figure}[\FBwidth]
{\caption{The typical bond of a vitrimer has to resist thermal fluctuations, but it has to allow swap if a triplet is established.}\label{fig2}}
{\includegraphics[scale=0.43]{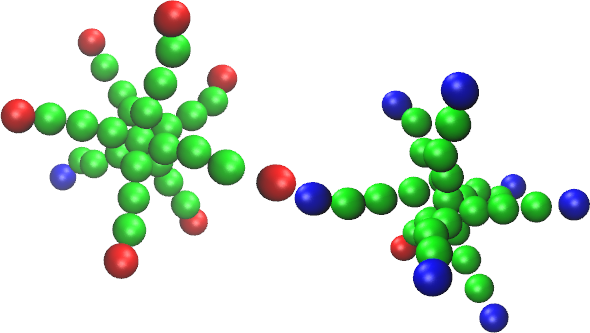}}
\end{figure}
The potential that we want to model between two active sites (i.e. a blue and a red bead in Fig.\ref{fig2}) has to be steep and strong in order to capture the strength of ester~\citep{capelot2012metal}, amide~\cite{denissen2015vinylogous}, alkene~\cite{lu2012olefin} or any bond that can produce vitrimers.
We base our model on a generalized Lennard-Jones (gLJ)
\begin{equation}
v^{\left( 2b \right)}_{ij}(\vec{r}_{ij})=v^{\left( gLJ \right)}_{ij}(\vec{r}_{ij})=
\begin{cases}
& 4 \epsilon \left[ \left( \dfrac{ \sigma}{r_{ij}} \right) ^{20}- \left( \dfrac{ \sigma}{r_{ij}} \right)^{10} \right] \qquad r_{ij}<r_\mathrm{cut}\\
& 0 \qquad \qquad \qquad \qquad \qquad \qquad \;\; r_{ij}\ge r_\mathrm{cut}
\end{cases}
\end{equation}
where we set $\epsilon=1$, $\sigma=1$ and at $r_\mathrm{cut}=2.5\sigma$. This is the attraction we impose between active sites to assemble the network. With just this attraction the system would inevitably clusterize, driven by multiple bond formation. 
We both prevent clusterization and provide a way to model the swapping processes simply combining that gLJ with a three body potential based on the same gLJ: 
\begin{equation}
V^{\left( 3b \right)}=\lambda \epsilon \sum_{ijk} \hat{v}^{ \left( 2b \right)}_{ij}(\vec{r}_{ij}) \cdot \hat{v}^{ \left( 2b \right)}_{ik}(\vec{r}_{ij})
\label{3body}
\end{equation}
where $\hat{v}^{ \left( 2b \right)}_{ij}$ is an auxiliary potential defined as:
\begin{equation}
\hat{v}^{ \left( 2b \right)}_{ij}(\vec{r}_{ij}) = 
\begin{cases}
& 1 \qquad \qquad \qquad  \;\,r_{ij}\le r_{min}\\
& - \dfrac{v^{ \left( 2b \right)}_{ij}(\vec{r}_{ij})}{\epsilon} \qquad r_{ij} > r_{min}\\
\end{cases}
\end{equation}
This solution is not only an elegant and smooth way to model a swap event, but it is also relatively cheap~\cite{sciortino2017three}. In fact we only have to combine the standard two body potentials to get the additional three body one, thus no extra computation is required.

It is also more natural to reproduce bond swap in this way compared to a Monte Carlo move, because the three body potential is structured such that a bond which is pulled is more likely to swap. To achieve the same result in MC we should force it using external tuning.

The three body term itself is noticeable only while a bond is swapping (i.e. when three ending beads are within the gLJ cutoff) otherwise it is zero and hence the system is characterized by two body terms only. So the only states that feel the presence of this additional potential are transient triplet states. Our assumption is that during a swap event no more than three sites are involved.

It is of fundamental importance to take those triplet states into account while evaluating thermodynamic quantities, because they deeply characterize the system (in fact without $V^{\left( 3b \right)}$ we should expect clusterization).
Then, in order to estimate the stress relaxation, we have to add those triplet terms in the pressure tensor.
To do so we can not rely on the standard virial approach~\cite{hansen1990theory} on which MD software are based, because one of its assumption is the pairwise interaction. 

In the following, we will derive a general expression for the isotropic pressure at equilibrium that includes the triplet contributions. Lastly we will generalize that expression
to its tensorial shape introducing the stress tensor as $\sigma_{\alpha\beta}= -P_{\alpha\beta}$. In this way we are able to measure the stress relaxation as discussed in the main text.

\subsection{Isotropic pressure}
It is possible to evaluate the average pressure in the canonical ensemble at fixed number of molecules $N$, volume $V$ and temperature $T$ using its definition
\begin{equation}
P \equiv -\left( \dfrac{\partial F}{\partial V}\right)_{T,N}
\label{pdef}
\end{equation}
and expressing the free energy $F$ from the partition function $Q$
\begin{equation}
F= -k_b T \,\mathrm{ log}\left(Q\right)
\end{equation}
then (\ref{pdef}) becomes
\begin{equation}
\beta P =\left( \dfrac{\partial Q}{\partial V}\right)_{T,N}
\label{bp}
\end{equation}
The partition function itself is defined as 
\begin{equation}
Q(N,V,T)\equiv\dfrac{1}{N!h^{3N}}\int d \{ r_i\} d \{ p_i \} \; \exp\left(-\beta H\right)
\end{equation}
where $H$ is the Hamiltonian that describes our system. It is possible to integrate analytically the momenta because their only dependence in the Hamiltonian is through the kinetic energy and that produces a gaussian integral. So we get
\begin{equation}
Q(N,V,T)=\dfrac{1}{N!\lambda_T ^{3N}}\int d \{ r_i\} \; \exp\left(-\beta U ( \{ r_i\} ) \right)
\label{part}
\end{equation} 
where the thermal wavelength corresponds to $\lambda_T=h/\sqrt{2\pi mk_BT}$. 

To get the pressure we have to take a derivative with respect to $V$ but the volume dependence is not only in the potential $U$, but also in the integration boundaries for the positions $\{r_i\}$. A solution is to use this new set of scaled variables
\begin{equation}
\xi_i \equiv \dfrac{r_i}{L}
\end{equation}
defined in this way to isolate the $V$ dependence from the integration domain. In fact (\ref{part}) becomes now
\begin{equation}
Q(N,V,T)=\dfrac{V^N}{N!\lambda_T ^{3N}}\int_0^1 d \{ \xi_i\} \; \mathrm{exp}\left(-\beta U ( \{ \xi_i\}, V ) \right)
\end{equation}
and we can finally derive the right side of (\ref{bp})
\begin{align}
\beta P 
& = \dfrac{1}{Q} \dfrac{\partial Q}{\partial V} \\
& = \dfrac{N}{V} - \beta \left< \dfrac{\partial U}{\partial V}  \right>
\end{align}
were the bracket $\left< . \right>$ denotes the ensemble average. If at this point we separate the two and three body potentials we get that
\begin{align}
\beta P 
& = \dfrac{N}{V} - \beta \left< \dfrac{\partial v^{(\mathrm{2b})}}{\partial V}  \right> - \beta \left< \dfrac{\partial v^{(\mathrm{3b})}}{\partial V}  \right> \\
& = \beta P^{(\mathrm{2b})} + \beta P^{(\mathrm{3b})} 
\end{align}
$P^{(\mathrm{2b})}$ is the standard pressure tensor in virial-2body approximation, which is widely discussed in any soft matter textbook~\cite{hansen1990theory,barrat2003basic}.
Instead $P^{(\mathrm{3b})}$, which is usually assumed to be zero, now requires a more detailed evaluation.

\emph{Factorizable 3 body potential - }
The three body potential in (\ref{3body}) is not a general function of three variables. Its three body dependence is only through
the sum of products of two variables at the time. Hence, there are no terms like $x_ix_jx_k$.
This means that its derivatives can be simplified further dividing each terms of the sum into two:
\def\vhbb{\hat{v}^\mathrm{(2b)}}
\begin{align}
\label{eq:isop}
\dfrac{\partial v^{(\mathrm{3b})}}{\partial V} 
& = \dfrac{1}{3L^2} \dfrac{\partial v^{(\mathrm{3b})}}{\partial L} \\
& = \dfrac{1}{3L^2} \dfrac{\partial }{\partial L} \left[ \sum_{ijk}^* \lambda \epsilon \vhbb_{ij} \cdot \vhbb_{ik} \right] \\
&  = \dfrac{\lambda \epsilon}{3L^2}     \sum_{ijk}^* \left[   \dfrac{\partial \vhbb_{ij} }{\partial L}  \cdot\vhbb_{ik} + \vhbb_{ij} \cdot \dfrac{\partial \vhbb_{ik}}{\partial L} \right]
\end{align}
With $\sum_{ijk}^*$ we mean that the summation has to be limited to the distinct triplets to avoid over counting. 
The derivatives of $\hat{v}$ are
\begin{equation}
\label{eq:comp}
\dfrac{\partial \vhbb(r_{ab})}{\partial L} = \dfrac{\partial \vhbb(r_{ab})}{\partial x_{ab}} \dfrac{\partial x_{ab}}{\partial L} + \dfrac{\partial \vhbb(r_{ab})}{\partial y_{ab}} \dfrac{\partial y_{ab}}{\partial L} + \dfrac{\partial \vhbb(r_{ab})}{\partial z_{ab}} \dfrac{\partial z_{ab}}{\partial L} 
\end{equation}
Due to the fact that $\vec{r}_{ab} \equiv L \vec{\xi}_{ab} $ we have for each of the component of (18)
\begin{equation}
\dfrac{\partial r_{\alpha}}{\partial L}= \xi_{\alpha} = \dfrac{r_{\alpha}}{ L}
\end{equation}
and then
\begin{equation}
\dfrac{\partial \vhbb(r_{ab})}{\partial L} = - \vec{F_{ab}} \cdot \dfrac{\vec{r_{ab}}}{L}
\end{equation}
Substituting (20) in (17) follows that
\begin{equation}
\dfrac{\partial v^{(\mathrm{3b})}}{\partial V} =-\dfrac{\lambda \epsilon}{3 L^3} \sum_{ijk}^* \left[   \vec{F}_{ij} \cdot \vec{r}_{ij} \; \vhbb_{ik} + \vhbb_{ij} \; \vec{F}_{ik} \cdot \vec{r}_{ik} \right]
\end{equation}
and then the total pressure
\begin{equation}
\label{eq:p}
\beta P = \beta P^{(\mathrm{2b})}_\mathrm{virial} + \beta \dfrac{\lambda \epsilon}{3 L^3} \sum_{ijk}^* \left[   \vec{F}_{ij} \cdot \vec{r}_{ij} \; \vhbb_{ik} + \vhbb_{ij} \; \vec{F}_{ik} \cdot \vec{r}_{ik} \right] 
\end{equation}
The main advantages of our 3 body potential is still shining in this formula in fact we already know all $\vec{F}_{ij}$ because we had to evaluate them for the dynamics of the system. So this expression is computationally cheap.

\subsection{Pressure and Stress tensor}
A formal expression for the stress tensor in the presence of our 3-body factorizable interaction can be derived in the framework of elasticity theory~\cite{landau1986theory}. 
A general deformation of a solid body can be defined as a transformation for its points $ \textbf{r}_\alpha' \equiv \Lambda_ \alpha ^{\;\,\beta} \textbf{r} _\beta $, where $\Lambda_{\alpha\beta}$ is the deformation tensor and $\alpha$ and $\beta$ represent the components $x,y,z$ in Einstein notation. It follows that any point is displaced by $\textbf{u}\equiv{\textbf{r}' }-\textbf{r} $ where we introduced the displacement field $\textbf{u}$ .
When an object is deformed the vector joining the same two points, called $dl$, is deformed as well.
Following Landau, in hypothesis of small deformation, we get that 
\begin{align}
(dl')^2  =  dl^2 +2u_{ik}dx_idx_k
\end{align}
$u_{ik}$ is an important quantity called the linearized strain tensor and defined by:
\begin{equation}
u_{ik}\equiv\dfrac{1}{2}\left(\dfrac{\partial u_i}{\partial x_k}+\dfrac{\partial u_k}{\partial x_i} \right)
\end{equation}
From its definition it is clear that it is symmetrical. It then means that it can be diagonalized at any point, i.e. we can choose co-ordinates axes in such a way that only $u_{ii}$ components of the tensors are non-zero (this will come in handy later on).

When a deformation occurs the arrangement of the molecules changes, hence thermal and mechanical equilibrium are broken. As a response, some portions of the body are going to have a resultant force different from zero that would like to rearrange the body in a new equilibrium. These internal forces produce the so called internal stress $\sigma_{ij}$. It is possible to evaluate those forces~\cite{landau1986theory} and determine them from the work done by the deformation $\delta W = - \sigma_{ik} \delta u_{ik}$. Its contribute to the  internal energy is $dE=TdS+\sigma_{ik} d u_{ik}$. Assuming instantaneous equilibrium we can introduce a free energy  $F=E-TS$ that for the deformed solid becomes
\begin{equation}
dF=-SdT+\sigma_{ik} d u_{ik}
\end{equation}
This allows us to get an expression for the stress tensor
\begin{equation}
\label{eq:str0}
\sigma_{ik} = \left( \dfrac{\partial F}{\partial u_{ik}} \right)_T
\end{equation}

\emph{Diagonal terms - }
For the diagonal terms $k=i$, the strain tensor and the stress are
\begin{equation}
u_{ii}=\dfrac{\partial u_i}{\partial x_i}=\Lambda_{ii}
\end{equation}
\begin{align}
\Rightarrow\sigma_{ii} 
& = \dfrac{\partial F}{\partial u_{ii}} \\
& = \dfrac{\partial F}{\partial \Lambda_{ii}} \\
& = \dfrac{\partial F}{\partial x'_i}  \dfrac{\partial x'_i}{\partial \Lambda_{ii}}  + \dfrac{\partial F}{\partial V'}\dfrac{\partial V'}{\partial \Lambda_{ii}}\\
& = \dfrac{\partial F}{\partial x'_i} x_i + \dfrac{\partial F}{\partial V'} V \\
& \approx \dfrac{\partial F}{\partial x'_i} x'_i  + \dfrac{\partial F}{\partial V} V
\end{align}
 to handle the derivative of the volume $V$ we used the fact that $V'= V + \Lambda_{ii} V + o \left( \Lambda^2 \right)$, then assumed small deformation, hence $\Lambda <<1$ and $x\approx x'$.  
So we get for the the $i=x$ component
\begin{equation}
\label{eq:a}
\sigma_{xx}=\dfrac{1}{V}\left(\dfrac{\partial F}{\partial x}x  + \dfrac{\partial F}{\partial V} V \right)
\end{equation}
where we normalized the stress dividing it by the entire volume $V$ of the system.
The stress tensor can be rewritten in this form
\begin{align}
\label{eq:str2}
\sigma_{ij} &= -P\delta_{ij}+\sigma^d_{ij}\\
&\equiv -p_{ij}
\end{align}
where $p_{ij}$ is the pressure tensor, $P=\frac{1}{3}\mathrm{tr}\left(p_{ij}\right)$ is the isotropic pressure and $\sigma^d_{ij}$ is the out of equilibrium dynamical stress. At equilibrium in isotropic condition, the relation states that $\sigma_{ii}=P=p_{jj}$ for any choice of $i$ and $j$ , as it can be easily verified comparing eq.\ref{eq:a} with eq.\ref{eq:p}~\cite{factor3}.

\begin{figure}
\floatbox[{\capbeside\thisfloatsetup{capbesideposition={right,center},capbesidewidth=5.5cm}}]{figure}[\FBwidth]
{\caption{A solid (black) after a deformation (red) defined by $x'=x+\Lambda z$ and $ z'=z+\Lambda x$. The same transformation can be rotated such that it becomes a single component deformation, changing the prospective from left to right. With this approach $x'=x+2\Lambda z$ but then in eq. (\ref{eq:aa}) we have to take the derivative only for a single direction.}\label{fig1}}
{\includegraphics[scale=0.43]{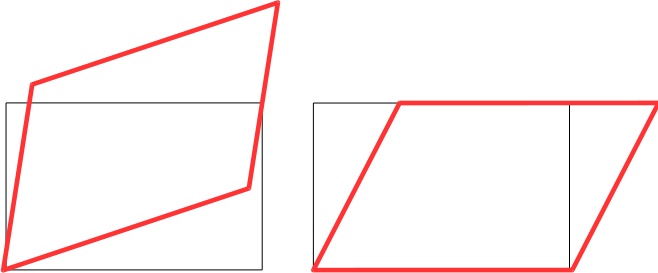}}
\end{figure}
\emph{Off-diagonal components - }
It is also possible to get the off diagonal components of the stress tensor. To do so, we start assuming symmetrical strain $\Lambda_{ij}=\Lambda_{ji}\equiv \Lambda$, then the off diagonal strains are:
\begin{equation}
u_{ij}=\dfrac{1}{2}\left(\dfrac{\partial u_i}{\partial x_j}+\dfrac{\partial u_j}{\partial x_i}\right)=\Lambda
\end{equation}
and then the stress
\begin{align}
\label{eq:2}
\sigma_{ij} 
& =  \dfrac{\partial F}{\partial u_{ij}} \\
\label{eq:aa}
& = \dfrac{\partial F}{\partial r'}\dfrac{\partial r'}{\partial \Lambda} + \dfrac{\partial F}{\partial V'}\dfrac{\partial V'}{\partial \Lambda}  
\end{align}
at this point we could evaluate the derivative of the free energy with respect to the strain parameter $\Lambda$, but it is easier to follow the argument sketched in Fig. \ref{fig1}. 
We can see that every off diagonal strain (i.e. one where we mix two components), can be expressed in a simpler way using the rotation depicted in the figure. In this way the $\Lambda$ dependence is on a single coordinate, hence it is easier to evaluate the derivative. After the rotation eq. (\ref{eq:aa}) becomes
\begin{align}
\sigma_{ij}  
& = \dfrac{\partial F}{\partial x_i}\dfrac{\partial x_i}{\partial \Lambda} + \dfrac{\partial F}{\partial V'}\dfrac{\partial V'}{\partial \Lambda}  \\
& = \dfrac{\partial F}{\partial x_i}x_j + \dfrac{\partial F}{\partial V'}\dfrac{\partial V'}{\partial \Lambda}  
\end{align}

Using the same tricks that we used for the pressure, we can decompose the stress tensor in an ideal gas term, a two body virial like term and a three body term that we evaluate again using factorization. The final expression is then:
\begin{equation}
\label{eq:3}
\sigma_{\alpha\beta}=-P_{\alpha\beta,\mathrm{virial}}^{(\mathrm{2b})}-\dfrac{\lambda\epsilon}{V} \sum_{ijk}^* \left[ 
   \left( \vec{F}_{ij}(r_{ij})\right)_{\alpha}  \left(\vec{r}_{ij}\right)_{\beta}\; \hat{v}^{ \left( 2b \right)}_{ik} 
   + \hat{v}^{ \left( 2b \right)}_{ij} \left( \vec{F}_{ik}(r_{ik})\right)_{\alpha}  \left(\vec{r}_{ik}\right)_{\beta} \right]   
\end{equation}


\begin{thebibliography}{38}
\bibitem [{tau()}]{tauliq_mes}%
  \BibitemOpen
  \href@noop {} {}\bibinfo {note} {We use the value $\tauliq\approx1\,\mu$s
  which is the fastest relaxation time that we get along the protocol. Using
  instead $\tauliq(\phi)$ would provide the same correlation for all of the
  liquid states.}\BibitemShut {Stop}%
\bibitem [{fac()}]{factor3}%
  \BibitemOpen
  \href@noop {} {}\bibinfo {note} {Keep in mind the assumption of isotropy that
  produces a factor 3 in eq.\ref{eq:p}, making the two expressions
  identical.}\BibitemShut {Stop}%
\end{thebibliography}
\end{document}